# Engineering the spin couplings in atomically crafted spin chains on an elemental superconductor

A. Kamlapure*, L. Cornils, J. Wiebe and R. Wiesendanger

*Department of Physics, University of Hamburg, Jungiusstrasse 9-11, D-20355 Hamburg, Germany*

*corresponding author: akamlapu@physnet.uni-hamburg.de

**Magnetic atoms on a superconductor give rise to Yu-Shiba-Rusinov (YSR) states within the superconducting energy gap. A spin chain of magnetic adatoms on an s-wave superconductor may lead to topological superconductivity accompanied by the emergence of Majorana modes at the chain ends. For their usage in quantum computation, it is a prerequisite to artificially assemble the chains and control the exchange couplings between the spins in the chain and in the substrate. Here, using a scanning tunneling microscope tip, we demonstrate engineering of the energy levels of the YSR states by placing interstitial Fe atoms in close proximity to adsorbed Fe atoms on an oxidized Ta surface. Based on this prototype platform, we show that the interaction within a long chain can be strengthened by linking the adsorbed Fe atoms with the interstitial ones. Our work adds an important step towards the controlled design and manipulation of Majorana end states.**

Majorana particles[1] have been proposed as key elements for topological quantum computation[2–5] due to their unique statistical properties. One of the systems for realizing Majorana modes in condensed matter is a spin chain on a superconductor[6–9]. The recent investigation of self-assembled ferromagnetic Fe chains on superconducting Pb, featuring strong spin-orbit coupling, triggered enhanced interest in the possible realization of Majorana modes at the ends of such chains[10–15]. So far, this platform for Majorana physics has been tested almost exclusively for the substrate material Pb hampering the controlled assembly of magnetic atoms into chains or more complex networks which are ultimately needed for braiding of Majorana modes and their usage in fault-tolerant quantum computation[3–5]. Moreover, the required YSR[16–18] band formation crucially depends on the exchange couplings between the spins within the chain[19–21] and between the spins and the conduction electrons of the host material[22,23]. Therefore, it is desirable to artificially construct such spin chains with full control over all the couplings in order to drive the system into desired topological phases[24].

To this end, we explore the superconducting substrate of a $(3 \times 3)$ oxygen reconstructed Ta(100) surface (named Ta(100)-$(3 \times 3)$O in the following) decorated with Fe atoms[23]. We first show that a spin chain constructed on the bare oxidized Ta(100) surface shows negligible interaction between nearest-neighbor Fe atoms. In the second step, we introduce Fe atoms in the interstitial sites in close proximity to the Fe adatoms using the tip of a scanning tunneling microscope (STM) and demonstrate the control over the energy levels of the YSR states. Finally we extend the method to build long chains of Fe adatoms that are interacting via interstitial Fe atoms and demonstrate spin coupling within the chain.

**Results**

**Assembly and investigation of chains of Fe adatoms.** As seen in the topographic STM image of the substrate (Fig. 1a), the regular network of oxygen atoms is imaged as a network of depression lines separating circular and cross-shaped plaquettes of $(3 \times 3)$ Ta atoms[23]. After low-temperature Fe deposition we see a statistical distribution of Fe adatoms with different apparent heights showing different spectroscopic features. In particular, one type of adatoms reveals a YSR state at a binding energy which varies for adsorption on a locally different substrate environment[23]. We were able to use vertical atom manipulation [25,26] to construct arrays of Fe adatoms positioned on the centers of neighboring cross-shaped plaquettes (Fig. 1 b-d, f; see Methods Section). Figure 1b-d show a single manipulated Fe adatom, a manipulated pair and a manipulated three-atom chain. Differential tunneling conductance (dI/dV) spectra taken with a superconducting tip on the manipulated atoms of the three structures (Fig. 1e) show a pair of peaks at energies close to the gap edge indicative of a YSR state with a relatively large binding energy ($E_b \sim \Delta$)[23]. Note, that the YSR state neither does change with the number of atoms in the chain nor with the number of neighbors, indicating negligible interaction amongst the atoms on neighboring plaquettes[20,21]. This holds for a long chain built from 63 atoms (Fig. 1f, Supplementary Fig. 1) where we see similar spectroscopic features along the chain as well as at the ends.

In order to verify the negligible coupling between neighboring Fe atoms in the chains, we used the following procedure. Applying a small bias pulse (typically ±500mV) to a manipulated Fe adatom, its electronic configuration can be switched between two states with strikingly different spectroscopic signature (Fig. 2a-d). For the original case discussed above (Fig. 2a) the spectrum shows the YSR states in the gap (Fig. 2c) and a resonance with a full width at half maximum on the order of 50 mV indicating a partial Kondo screening of the magnetic moment of the Fe atom (Fig. 2 d)[22]. In the following, we denominate this state of the Fe adatom as the "YSR-on" state. After applying a positive voltage pulse (Fig. 2b) the spectrum on the Fe adatom completely changed to the spectroscopic signature of a substrate spectrum (Fig. 2c,d). This most probably indicates that the Fe adatom has completely lost its magnetic moment. We refer to the corresponding state as the "YSR-off" state. Note, that the process is reversible by the application of a voltage pulse of reversed bias polarity. We suppose that the voltage-pulse induced switching of the electronic and magnetic properties of the Fe adatom is possibly due to a switching between two different metastable states of the Fe atom on the Ta(100)- (3 × 3)O reconstruction[27,28]. Using these two different spectroscopic signatures, we can also identify the state of the corresponding Fe adatom in a dI/dV image recorded at $V = 50$ mV (insets of Fig. 2a, b) revealing a lower dI/dV signal in the YSR-on state as compared to the YSR-off state. Below, we will investigate the effect of the controlled switching of an Fe adatom within the chains between the magnetic (YSR-on) state and the non-magnetic (YSR-off) state on the YSR states of the neighboring atoms, which will enable us to conclude on the magnetic coupling between the atoms in the chain[20,21]. This is demonstrated in a four-atom chain (Fig. 2e-g). When an atom within the chain is switched from the YSR-on state (Fig. 2e) to the YSR-off state (Fig. 2f shown exemplarily for the second atom from the top), the dI/dV spectra on all other atoms within the chain look almost identical (Fig. 2g). We, therefore, conclude that chains built from adatoms on neighboring plaquettes reveal negligible interatomic Fe interaction preventing the formation of a YSR band, as expected by the rather large interatomic distance (~ 1nm) with a decoupling oxygen row in-between[23].

**Interstitial atom manipulation and creating Fe arrays.** In order to increase the coupling between the Fe adatoms, we devised a unique technique of manipulating Fe atoms into interstitial sites in the substrate in-between the Fe adatoms. Using the STM tip employing vertical manipulation (see Methods Section), an Fe atom is first dropped close to the target region onto an O row in-between two plaquettes (Fig. 3a,b). With the tip held above this Fe atom, the bias is then slowly increased (Fig. 3d) until at a typical bias voltage between 100 mV and 300 mV a sudden decrease in the z position of the tip is seen. This indicates that the Fe atom moved into an interstitial site in the middle between two plaquettes, as shown by the slightly increased apparent height of the corresponding O row (Fig. 3c). In the following, we name such an Fe atom as an interstitial Fe atom (IFA). The dI/dV spectra taken on the IFAs show one or two tiny replicas of the coherence peaks outside the gap (Supplementary Fig. 2) most probably due to spin excitations[29], indicating that the IFAs have a residual magnetic moment. Using this technique, we assembled adatoms with one, two and three IFAs in close proximity as shown by the white arrows in the STM images (Fig. 3e-h). The pictograms of various assemblies are shown in Figure 3i. The dI/dV spectra on the adatoms of each of these assemblies (Fig. 3i, in order to remove the effect of the superconducting tip, their numerical deconvolution extracted as described in the Methods section is illustrated in Fig. 3j) reveal that, with increasing number of IFAs, the most intense YSR state at positive bias voltage shifts from the gap edge towards the Fermi energy (A-C) and finally appears on the negative bias voltage side (D). This observation of a shift in the binding energy of the YSR state is consistent with a decreasing effective Kondo coupling of the Fe adatom with the substrate mediated through the increasing number of IFAs[22,23]. However, since the spectra indicate a nonzero magnetic moment of the IFA (Supplementary Fig. 2), the situation is more complex, as the Fe adatom is additionally exchange coupled to the IFA. Both, strong ferromagnetic (FM) and antiferromagnetic (AFM) exchange interactions, are known to renormalize the effective Kondo coupling[30–34]. Additionally, for the FM case, the dI/dV spectra are expected to show multiple YSR peaks while, for the AFM case, only one pair of shifted peaks is expected[19,21,35]. Indeed, a careful analysis of the spectrum corresponding to adatom B (Supplementary Fig. 3) shows additional small peaks indicating a small FM interaction between the adatom and the IFA for this particular case. Since the surface is theoretically shown to be metallic[36], we expect an RKKY type of interaction playing a significant role in the coupling between the IFA and the Fe adatom. However, we cannot rule out the possibility of a superexchange mechanism mediated through the oxygen atoms which are bridging the adatom and the IFA. A detailed theoretical treatment of the exchange mechanism is beyond the scope of the current paper.

**Spin coupling in pairs of Fe adatoms using one IFA.** Motivated by the results described above, we investigate the coupling of a pair of adatoms with one IFA in the center using the technique of the switching of the adatoms between their YSR-on and -off states (Fig. 4).

When both adatoms are in the YSR-on state (middle panel in Fig. 4a), they exhibit a YSR state at a distinct energy similar to that of the type B structure in Figure 3f, which is clearly different from that of the Fe pair without IFA (c.f. Fig. 1e). However, the bound state energy positions of the two adatoms are not exactly the same showing an asymmetric coupling of both adatoms in the pair to the IFA. The asymmetry could possibly result from either a slightly asymmetric position of the IFA between the two adatoms or an intrinsic variance in the structure of the plaquettes of (3 x 3) Ta atoms (Fig. 1a) which can result in a modified coupling of the IFA with the adatoms on top. When one of the two adatoms in the pair is switched into the YSR-off state (upper and lower panels in Fig. 4a,b) the YSR state of the remaining YSR-on adatom shifts towards the Fermi level. Note, that this result is strikingly different from that of the pairs and chains without IFAs (Fig. 2e-g) where no changes in the spectra were observed when switching one of the atoms into the YSR-off state. This clearly shows that the interaction between two neighboring adatoms can be increased via the IFA (see additional spectra on a pair with an IFA in Supplementary Fig. 4).

**Chains of Fe adatoms with interstitial Fe atoms.** We consequently extend the scheme described above to the formation of a chain of 13 adatoms interacting via IFAs. To this end, we initially created a chain of IFAs (Fig. 5a) followed by manipulating a chain of adatoms on top (Fig. 5b). The dI/dV spectra of all adatoms of this chain (solid lines of Fig. 5d) show the strongest YSR state well inside the gap, and a strong variation of the YSR binding energies along the chain. To check the coupling along the chain, the $6^{th}$ Fe atom from the top was switched from the YSR-on to the YSR-off state (Fig. 5c), and the according spectra are shown by dashed lines in Figure 5d. The switching has a most prominent effect on the $5^{th}$ adatom from the top (Fig. 5e,f). The deconvoluted spectra (Fig. 5f) clearly show that at least four prominent peaks are needed in order to correctly capture the details of the spectra in Figure 5e (see Supplementary Fig. 5). This implies that there is a considerable splitting of the YSR state which is strongly indicating FM couplings[20,21,35] between the adatoms in the chain. From the vertical green lines in Figure 5f it is evident that, when the nearest atom is switched from YSR-on to YSR-off state, the separation between the two peaks due to the splitting of YSR states is reduced. This is consistent with a reduction of the spin coupling between the two neighbors by switching one of them into a non-magnetic state. Similar measurements on the 6$^{th}$ atom (see Supplementary Fig. 6) also show the splitting of the YSR state. However in this case, when the nearest neighbor is switched into the YSR-off state, we do not see a change in the splitting within our energy resolution, but only a strong variation of the intensities of the peaks. While these results evidence an exchange interaction between the adatoms in the chain, we cannot rule out that switching the atom from YSR-on to YSR-off may imply a small structural change in the nearest neighbours of the switched atom, which might cause additional shifts in the YSR states[37]. Figure 5d, furthermore reveals that the switching of the 6$^{th}$ atom has a prominent influence only on the 5$^{th}$ atom, but not on the 7$^{th}$ atom, which is suggestive of an asymmetric coupling along the chain. Finally, similar switching experiments performed on all the other atoms of the chain (not shown) reveal distinct changes in the spectral features only in the nearest neighbors of the switched adatom, while the spectra away from this atom show little changes. Accordingly, we conclude that the IFAs indeed induce a short-range exchange coupling between the Fe adatoms in the chain. Note, that the short-range couplings in combination with the inherent asymmetry in the coupling between two adatoms in a pair hamper the formation of a coherent YSR band along the chain. We accordingly do not observe any signature of zero bias peaks at the chain ends in the present study. We, however, anticipate that stronger and long-range couplings might be possible by positioning additional IFAs aside the chain which could result in the formation of Majorana end modes. Therefore, our system provides an ideal playground for the tuning of various couplings via IFAs and additionally is a promising system to explore novel physics of spin-related phenomena in superconductor-spin chain hybrids.

**Conclusions**

In summary, we showed that the system of Fe adatoms on the Ta(100)-$(3 \times 3)$O surface provides a promising playground for crafting magnetic chains of various length coupled to an s-wave superconductor, and tuning the different couplings within the system by interstitial Fe atoms. We demonstrated a strengthening of the magnetic interactions along the chain via the interstitial Fe atoms and a manipulation of the binding energy of the YSR state. This method can be extended to more complex systems as, e.g., networks of coupled chains[38]. Our work is thus an important step towards the controlled realization and manipulation of topological superconductivity and Majorana end modes.

**Methods**

**Experimental methods.** All scanning tunneling microscopy (STM) and spectroscopy (STS) measurements were carried out in a custom built SPECS STM under ultra-high vacuum conditions and at the base temperature of 1.1 K using an electrochemically etched bulk Cr-tip[39] coated with tantalum[23]. Details of the surface cleaning as well as Fe atom deposition can be found elsewhere[23].

Prior to the atom manipulation to create various arrays of magnetic adatoms, the surface was cleaned (Fig. 1a) by picking up all the adatoms using vertical manipulation. To this end, the tip was initially held on the atom with typical stabilization parameters of V = 50 mV and I = 50 pA. With feedback off, a typical bias voltage pulse of ~ +1V was applied in order to transfer the atom to the tip. For the transfer of Fe atoms from the tip to the center of the plaquettes or in-between two plaquettes the feedback was switched off and the tip was lowered by ~300-600 pm and a negative bias voltage pulse of 0.3V-1V was applied in discrete steps until an Fe atom dropped, as seen in both the z position of the tip and the image acquired afterwards. We did not observe YSR states on the Fe atoms adsorbed on the circular plaquettes. Therefore, we created all the atomic arrays by adatoms positioned on the cross-shaped plaquettes. Once the desired atomic arrays were created, the tip was mechanically dipped into the Ta surface in order to coat the tip apex with a superconducting Ta cluster.

Differential conductance spectra (d$I$/d$V$) were measured using a standard lock-in technique[40] after stabilizing the tip at $V_{stab}$ and $I_{stab}$ with a typical modulation voltage of $V_{mod}$ = 20 μV ($f$ = 827 Hz).

**Numerical deconvolution of the spectra.** For the deconvolution of the spectra measured with a superconducting tip, initially the tip density of states $N_t$ is obtained by numerically fitting the substrate spectrum to the expression for the lock-in detected tunneling spectrum of a superconductor-insulator-superconductor junction, given by[40]

$$\frac{dI}{dV}(V) \propto \int_{-\pi/2}^{\pi/2} \sin\alpha \cdot I(V + \sqrt{2}V_{\text{mod}}\sin\alpha, T)d\alpha. \tag{1}$$

Here, $V_{\text{mod}}$ is the modulation voltage used for the lock-in measurements and

$$I(V,T) \propto \int_{-\infty}^{\infty} N_t(E) \cdot N_s(E+V) \cdot [f(E) - f(E+V)] \cdot dE, \tag{2}$$

where, $f(E)$ is the Fermi function and $N_{t(s)}$ is the BCS-Dynes density of states of the tip (sample) given by

$$N(E,\Gamma) = N_n(E_F) \cdot \Re\left[\frac{E + i\Gamma}{\sqrt{(E+i\Gamma)^2 - \Delta^2}}\right]. \tag{3}$$

We get the best fitting parameters for the tip with $\Delta_t = 0.5$ meV and $\Gamma_t = 0.04$ meV. Once the tip is characterized, we model the density of states for the YSR states as a sum of a gap and $n$ Lorentzian peaks given by[22],

$$N_s = \frac{1}{e^{\frac{\Delta_s - |E|}{\delta_s}} + 1} + \sum_{i=1}^{n} \frac{A_i}{1 + \left(\frac{E - E_i}{\gamma_i}\right)^2} \tag{4}$$

Here, $\Delta_s$, $\delta_s$, $A_i$, $E_i$ and $\gamma_i$ are the free parameters which are determined using nonlinear least square fitting.

**Data availability**

The authors declare that all the relevant data are included in the paper and its supplementary information files. Additional data are available from the corresponding author upon request.

**Competing interests**

The authors declare no competing interests.

**Acknowledgements**

The authors thank Maria Valentyuk, Roberto Mozara, Alexander I. Lichtenstein, Saurabh Pradhan and Jonas Fransson for fruitful discussions, and Dörte Langemann for help with the manipulation. This work has primarily been supported by the ERC Advanced Grant ASTONISH (No. 338802). L.C. and J.W. acknowledge support through the DFG priority programme SPP1666 (grant no. WI 3097/2).

**Author contributions**

All authors conceived and designed the experiments. A.K. and L.C. performed the experiments. A.K. analyzed the data. A.K. and J.W. wrote the manuscript. All authors discussed the results and commented on the manuscript.

# Figures

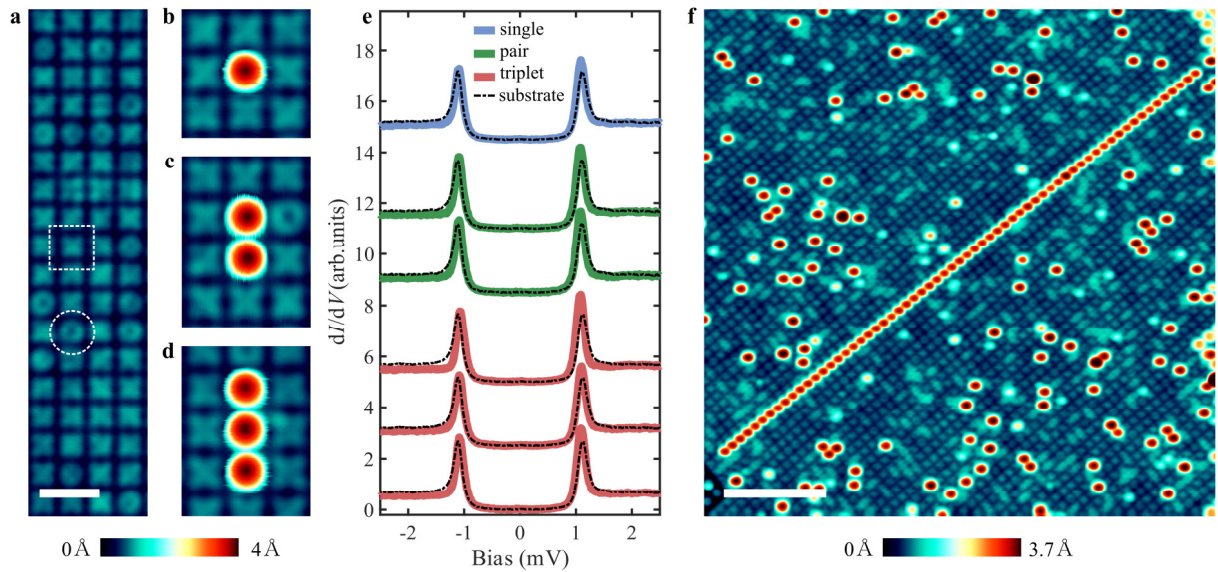

**Figure 1 | Spectroscopy of short chains of Fe adatoms. a,** Constant-current STM image of the typical surface of Ta (100)-O showing a regular array of cross shaped (dashed rectangle) and circular shaped (dashed circle) plaquettes. White scale bar is 2 nm. **b-d,** STM images showing a single Fe adatom, a pair, and a three-atom chain that were assembled using atom manipulation. **e,** d$I$/d$V$ spectra measured with a superconducting tip on top of the different chains of Fe adatoms shown in **b-d**. Each spectrum is plotted together with the spectrum of the substrate taken with the same tip apex. Spectra are shifted vertically and grouped for clarity (STS parameters: $V_{stab}$ = 2.5 mV, $I_{stab}$ = 100 pA, $V_{mod}$ = 20 μV, scan parameters: $V$ = 50 mV, $I$ = 100 pA). **f,** STM image of a long chain made of 63 Fe adatoms. White scale bar is 10 nm long ($V$ = 100 mV, $I$ = 100 pA).

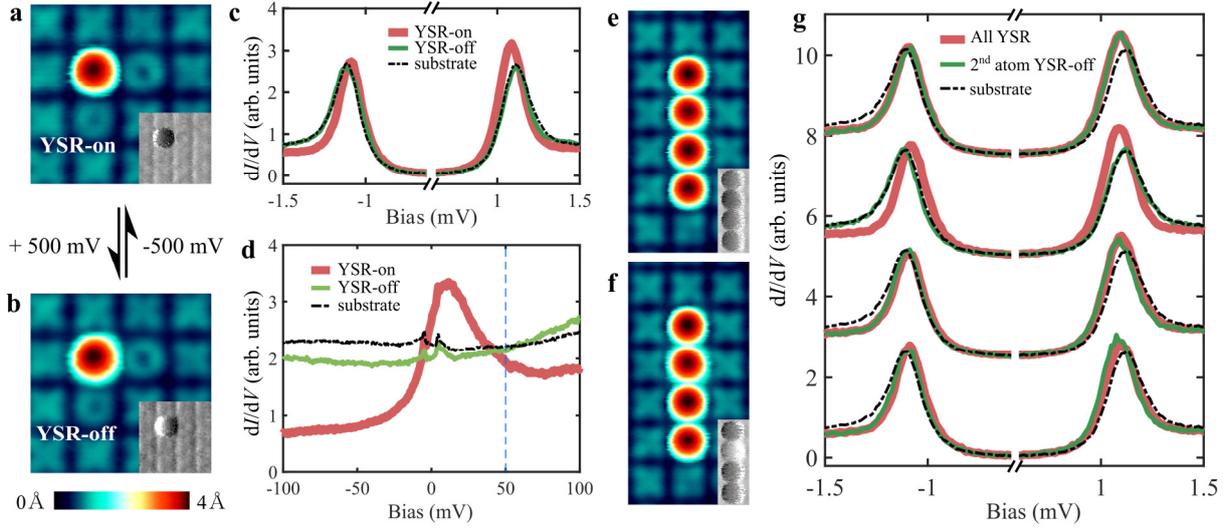

**Figure 2 | Switching between YSR-on and YSR-off state in a four-atom chain. a,b,** STM image showing a single Fe adatom in YSR-on and YSR-off state, respectively. The typical voltage pulse required to switch between the two states is indicated between the two panels. The inset in each panel shows the d$I$/d$V$ image at V = 50 mV. **c,d,** dI/dV spectra taken with a superconducting tip on the adatom for the two cases shown in **a** and **b,** plotted together with the substrate spectrum. **e,f,** STM image of a four-atom chain with the second atom in YSR-on and YSR-off state, respectively. Insets in each panel represent dI/dV images. **g,** d$I$/d$V$ spectra taken with a superconducting tip on each adatom of the four-atom chain for the two cases shown in **e** and **f,** plotted together with the substrate spectrum. The spectra are shifted vertically for each adatom for clarity. Scan parameters: $V$ = 50 mV, $I$ = 100 pA, $V_{mod}$ = 4 mV. Stabilization parameters for STS in **c** and **g**: $V_{stab}$ = 2.5 mV, $I_{stab}$ = 100 pA, $V_{mod}$ = 20 μV and those for **d**: $V_{stab}$ = 100 mV, $I_{stab}$ = 100 pA, $V_{mod}$ = 3 mV.

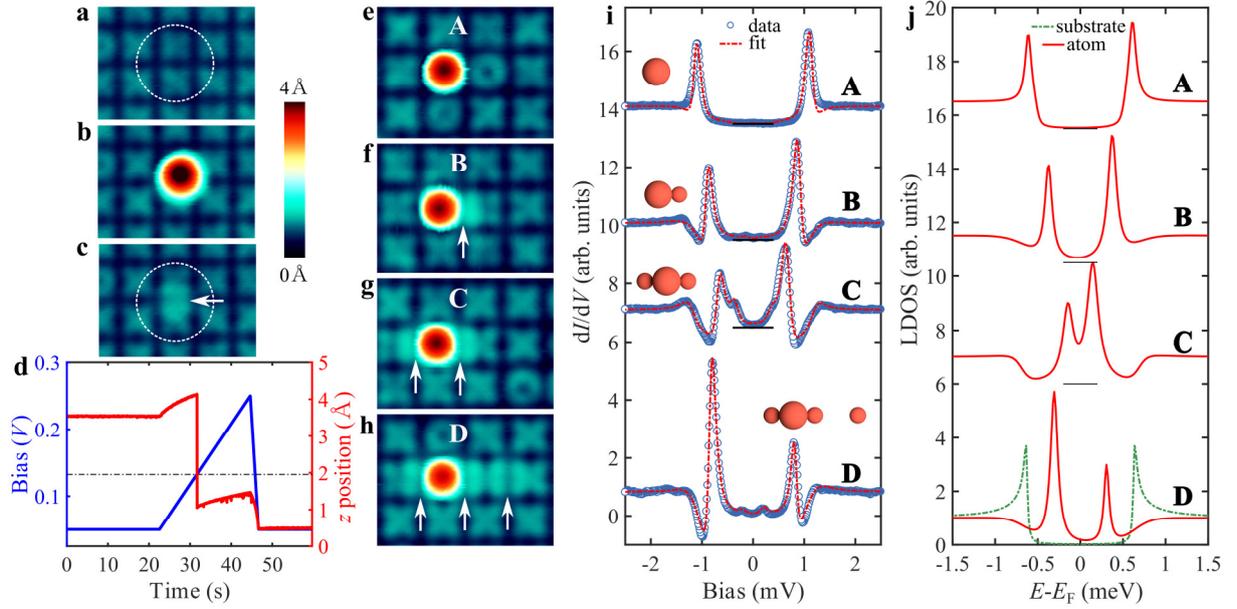

**Figure 3 | Spectroscopy on adatoms coupled to interstitial Fe atoms (IFAs). a,** STM image of the area before the construction of the IFA. Dashed circle represents the target point. **b,** STM image after vertical manipulation of an adatom to the target point. **c,** Image after introducing the IFA into the target site. **d,** Bias voltage and $z$ position of the tip as a function of time during the manipulation of the Fe from the position in **b** to the interstitial site in **c**. Horizontal dashed line indicates the bias value, $V = 130$ mV, at which the Fe adatom relaxes into the interstitial site. **e-h,** STM image of adatoms with no (A), one (B), two (C) and three (D) IFAs in close proximity. Here, the white arrows represent the positions of IFAs. **i,** $dI/dV$ spectra taken with a superconducting tip along with the fit using numerical deconvolution corresponding to the adatoms shown in **e-h**. Spectra are shifted vertically for clarity. The pictograms indicate the positions of adatoms (large spheres) and IFAs (small spheres). **j,** Numerically deconvoluted local density of states (LDOS) corresponding to the fits in (**i**). Substrate density of states is plotted as reference. Stabilization parameters for STS: $V_{stab} = 2.5$ mV, $I_{stab} = 100$ pA, $V_{mod} = 20$ µV and scan parameters: $V = 50$ mV, $I = 100$ pA.

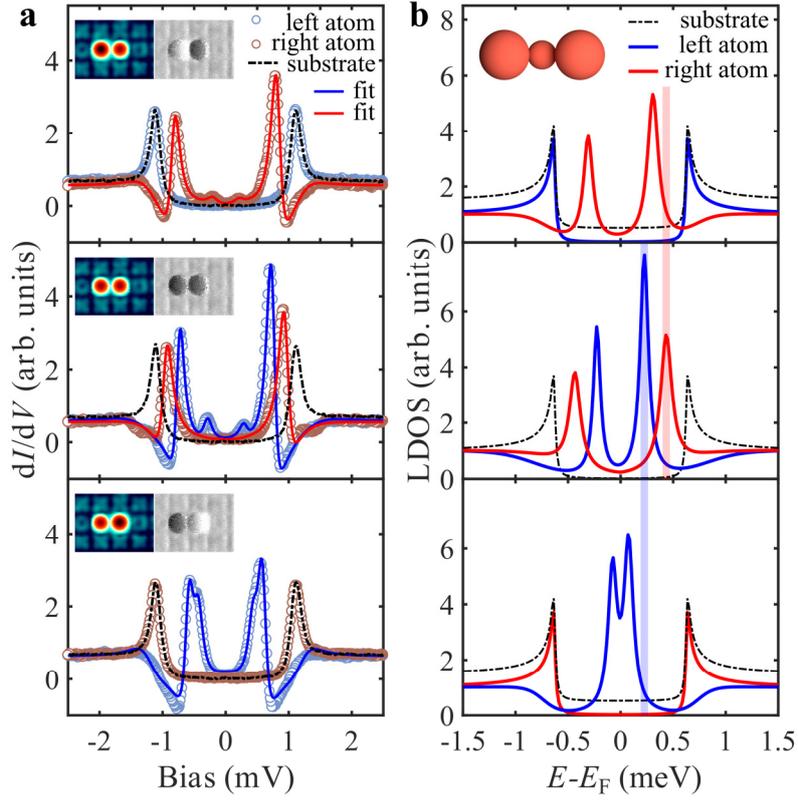

**Figure 4 | Spectroscopy on a pair with one IFA in the center. a,** d$I$/d$V$ spectra taken with a superconducting tip on the two adatoms plotted together with substrate spectrum for the three cases: (1) top panel: left atom in YSR-off state and right atom in YSR-on state, (2) middle panel: both atoms in YSR-on state, (3) bottom panel: left atom in YSR-on state and right atom in the YSR-off state. The solid curves in each panel represent a fit using the numerical deconvolution. Insets in each panel are the corresponding STM and d$I$/d$V$ images. **b,** The numerically deconvoluted LDOS corresponding to the fits in each panel in **a.** The black dashed curve corresponding to the substrate DOS is shown in each panel as a reference. Vertical lines are a guide to the eye showing shifts in the YSR state energy from the case (2) to the cases (1), (3). The pictogram in the upper panel shows the position of the Fe atoms in the assembly. Substrate DOSs in upper and lower panels are shifted vertically for clarity. Stabilization parameters for STS: $V_{stab}$ = 2.5 mV, $I_{stab}$ = 100 pA, $V_{mod}$ = 20 µV and scan parameters: $V$ = 50 mV, $I$ = 100 pA, $V_{mod}$ = 3 mV.

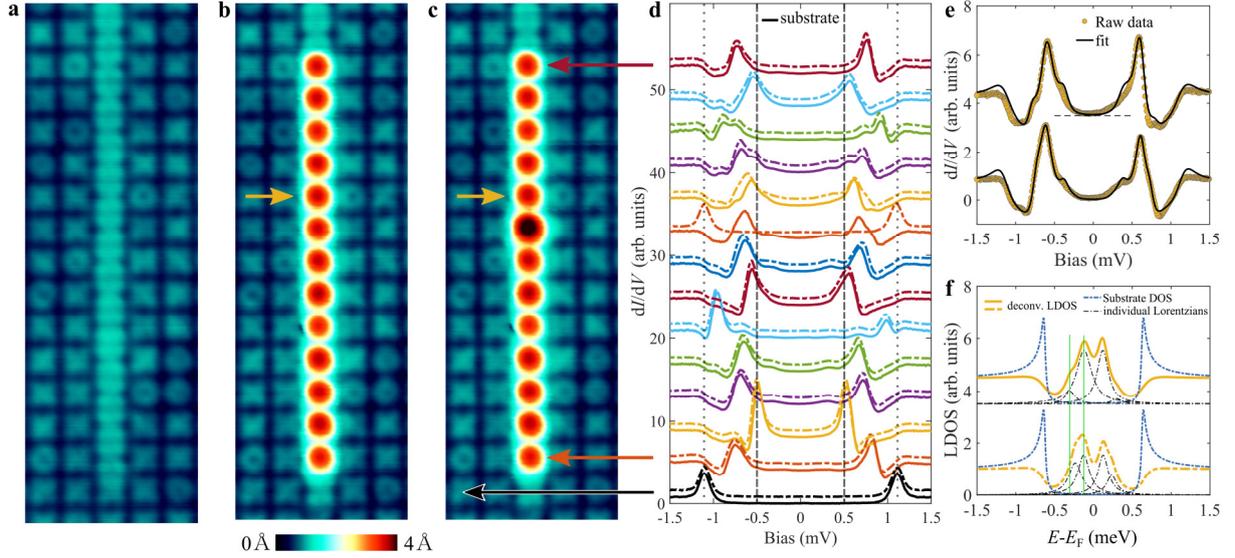

**Figure 5 | Spectroscopy on a chain of 13 atoms a,** STM image of manipulated IFAs before the construction of the chain. **b,c,** STM images of the chain with the 6$^{th}$ atom from the top in YSR-on and YSR-off state, respectively. The remaining atoms are in the YSR-on state for the two cases. **d,** d$I$/d$V$ spectra taken with a superconducting tip on each adatom of the chain corresponding to (**b**) (continuous curves) and (**c**) (dashed curve). Black curves at the bottom represent substrate spectra. The spectra are shifted vertically for clarity. The dashed (dotted) vertical lines indicate the sample Fermi level (sample coherence peak). **e,** d$I$/d$V$ spectra on the 5$^{th}$ atom from the top (same as orange curves in (**d**)), together with the fits from numerical deconvolution of the spectra in two different situations corresponding to (**b**) (top curves) and (**c**) (bottom curves). **f,** The numerically deconvoluted LDOS corresponding to the fits in panel (**e**). The substrate DOS is plotted as a reference. Vertical green lines in (**f**) are guides to the eye showing the peak position of the YSR states in the upper most curve. Stabilization parameters for STS: $V_{stab}$ = 2.5 mV, $I_{stab}$ = 100 pA, $V_{mod}$ = 20 μV and scan parameters: $V$ = 50 mV, $I$ = 50 pA.
18

# Supplementary Figures

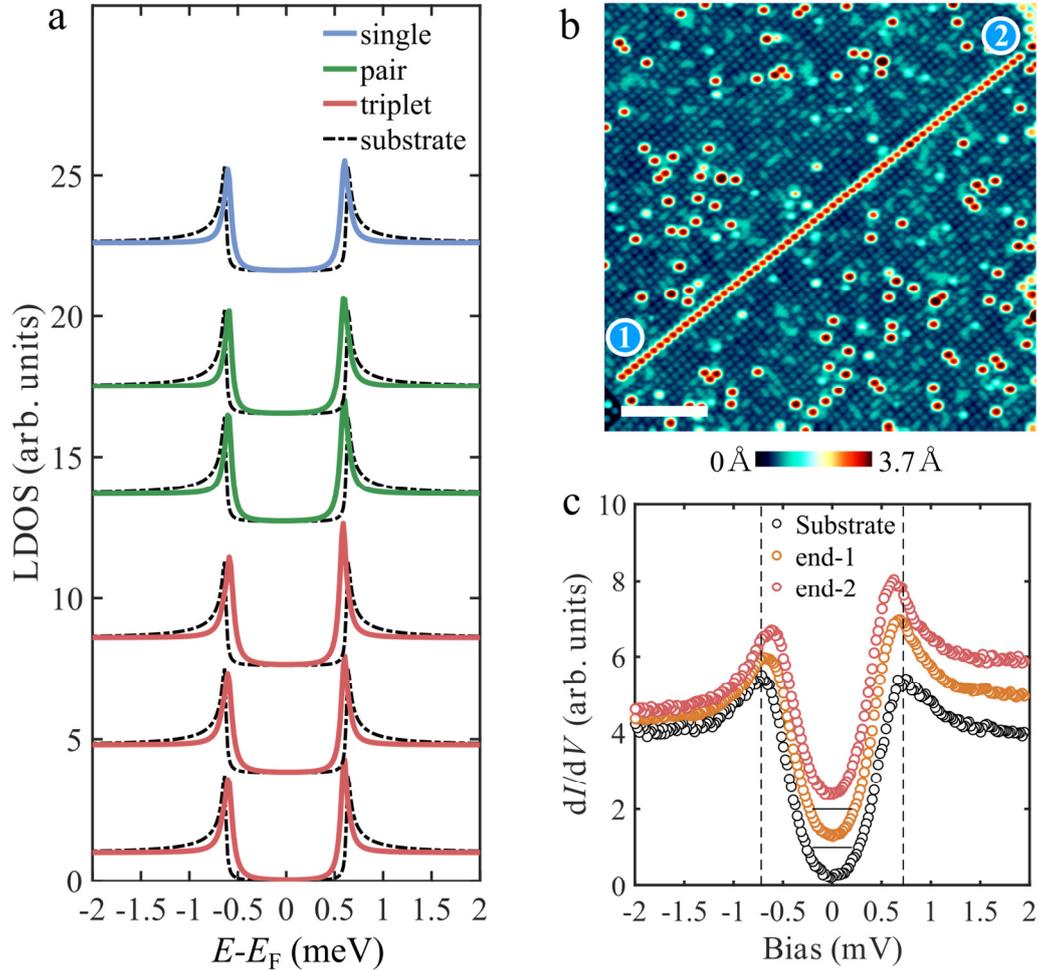

**Supplementary Figure 1 | Spectroscopy on different chains. a,** The numerically deconvoluted local density of states (LDOS) corresponding to the data in Fig. 1e of the main manuscript. The curves are shifted vertically and grouped for clarity. All the curves are plotted together with the substrate density of states as a reference. Resemblance of these curves indicates the negligible coupling between adatoms in the chains. **b,** Constant-current STM image of the long chain of 63 atoms (same as Fig. 1f in the main manuscript, $V = 100$ mV, $I = 100$ pA). Scale bar is 10 nm. **c,** d$I$/d$V$ spectra measured with a Cr-tip at the two ends of the chain (end-1 and end-2 shown in **b**) plotted along with the substrate spectrum. For clarity, the spectra are shifted vertically ($V_{stab} = 6$ mV, $I_{stab} = 300$ pA, $V_{mod} = 100$ μV). Spectra taken at the two ends show peaks due to YSR states close to the gap edge. This energetic position of the YSR state is identical to the ones in **a**, and Fig. 2c,g of the main manuscript, showing a negligible coupling along the chain without any prominent features in the spectra.

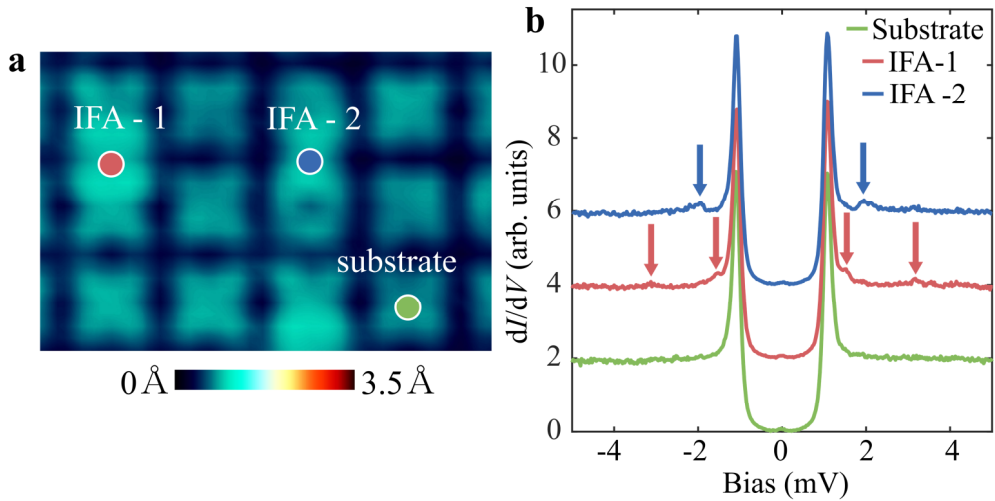

**Supplementary Figure 2 | Spectroscopy on the interstitial Fe atoms (IFA). a,** Constant-current STM image showing two IFAs which are manipulated in two different surroundings of the superstructures. **b,** d$I$/d$V$ spectra taken with a superconducting tip on the three locations shown in **a** ($V_{stab}$ = 2.5 mV, $I_{stab}$ = 300 pA, $V_{mod}$ = 20 µV). Spectra are shifted vertically for clarity. Both IFA-1 and IFA-2 show replicas of the coherence peaks (see arrows), possibly arising from a spin excitation[1]. However, the detailed features are different for the two cases, which implies slightly different positions of the IFAs in the interstitial sites. Moreover, the replicas might indicate the presence of a finite magnetic moment of the IFA.

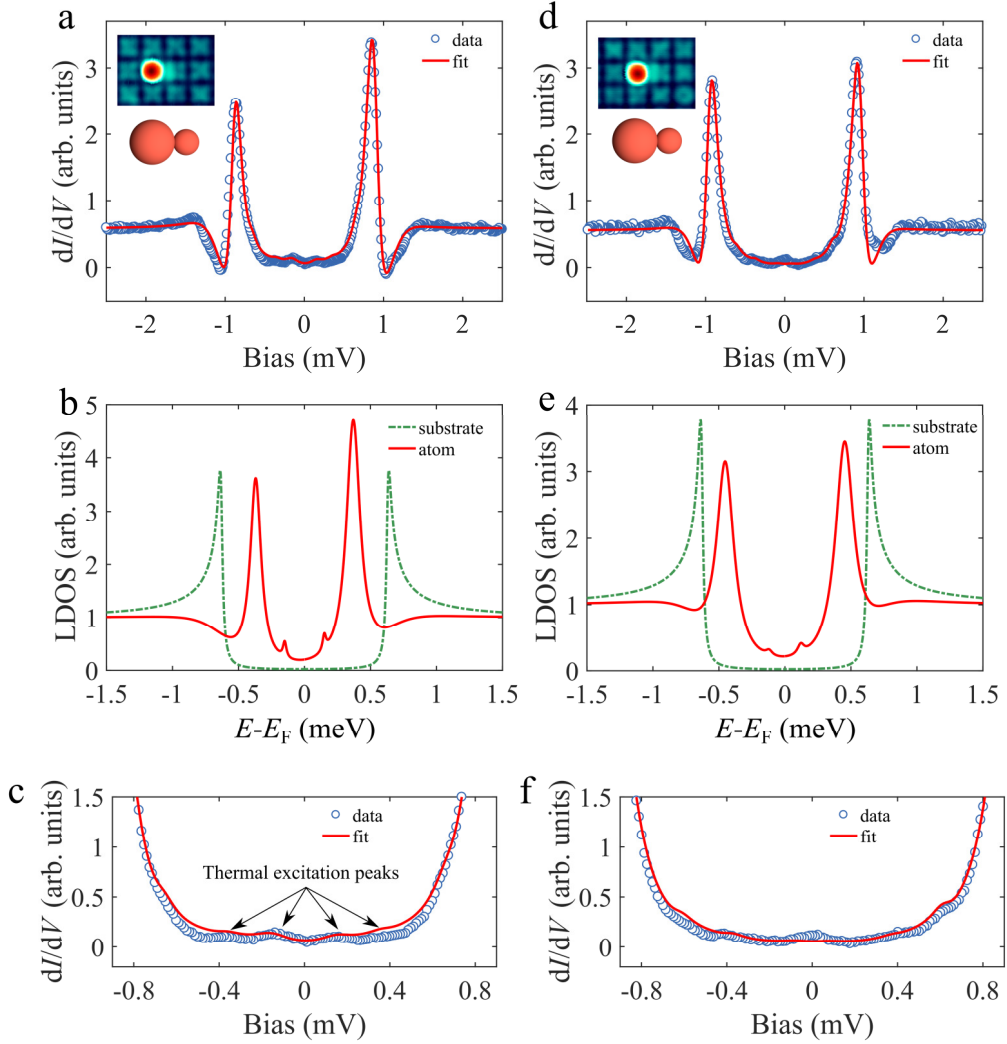

**Supplementary Figure 3 | Spectroscopy on an adatom with one IFA in the vicinity. a,** d$I$/d$V$ spectrum taken with a superconducting tip on adatom B (Fig. 3**f** of the main manuscript) plotted together with the fit using numerical deconvolution. The inset shows the STM image of the adatom. The pictogram indicates the positions of the adatom (large sphere) and the IFA (small sphere). **b,** The numerically deconvoluted local density of states (LDOS) on the adatom plotted together with the substrate density of states as a reference. To describe the extra features in the d$I$/d$V$ spectrum on adatom B (**a**), we need to use multiple YSR states to capture all the spectral features. For a symmetric magnetic dimer with ferromagnetic coupling, four symmetric peaks are expected[2]. In our case, we indeed see two additional peaks, but with a much smaller intensity than that of the main two peaks. **c,** The zoomed-in view of the same spectrum and the fit as in **a**. The fit captures details of the spectral features due to thermally occupied states within the superconducting energy gap of the tip, which are otherwise not prominently visible in **a**. **d-f**, Similar spectral analysis as in **a-c** on another adatom of type B. Again, we need to use multiple YSR states to capture the details of the additional spectral features inside the gap.

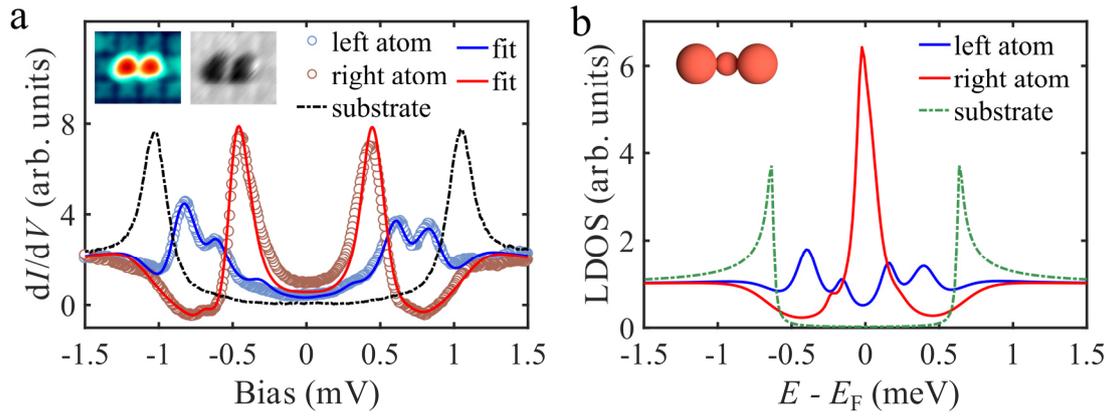

**Supplementary Figure 4 | Spectroscopy on a pair with one IFA in the center. a**, d$I$/d$V$ spectra taken with a superconducting tip on a pair of adatoms with one IFA in the middle, plotted together with the substrate spectrum. Here both of the adatoms are in the YSR-on state. The solid curves represent a fit using the numerical deconvolution. Insets in the panel are the STM image and the d$I$/d$V$ image. **b,** The numerically deconvoluted LDOS corresponding to the fits in (a). The substrate spectrum is shown as a reference. The pictogram indicates the positions of the adatoms (large spheres) and the IFA (small sphere). To describe the extra features in the d$I$/d$V$ spectrum on each of the atoms in the pair (**a**), we need to use multiple YSR states that capture all the spectral features. This, again, indicates a ferromagnetic coupling[2] to the IFA or between the two adatoms, similar to the structure discussed in Supplementary Figure 3.

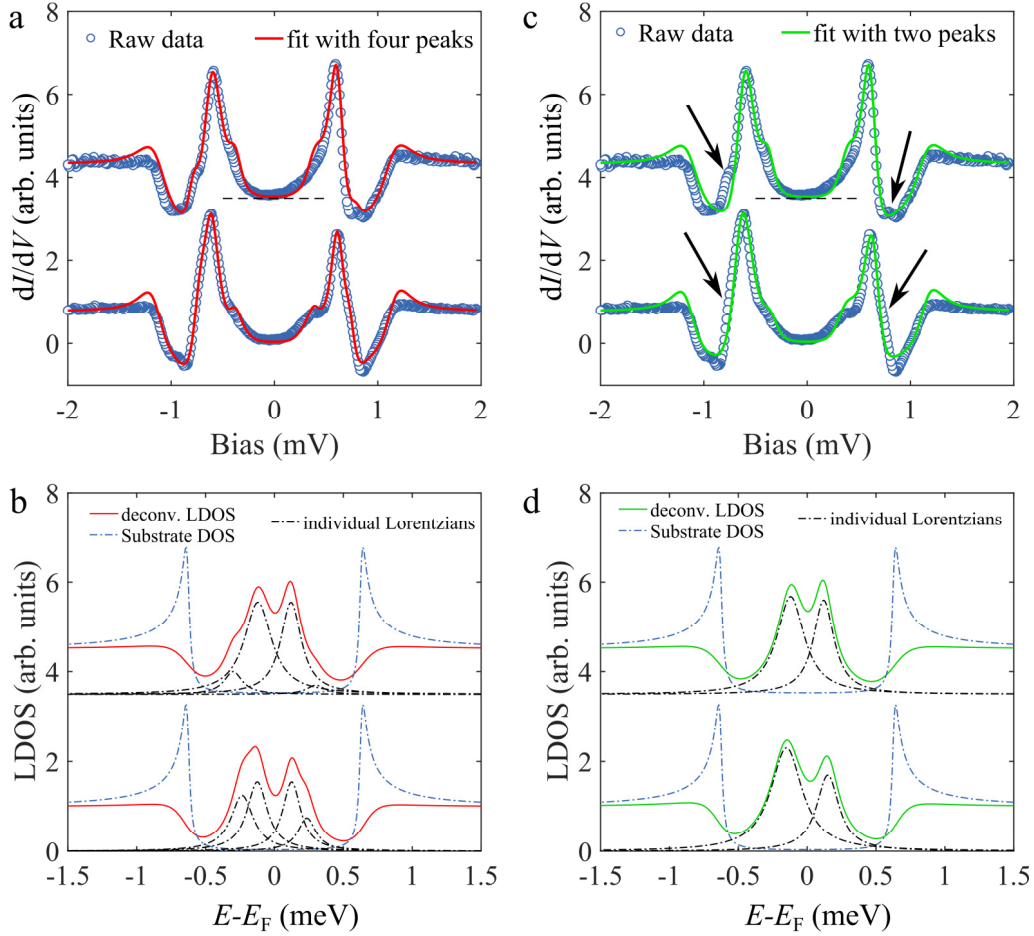

**Supplementary Figure 5 | Assessment of the fitting. a,** d$I$/d$V$ spectra (same as Fig. 5e of the main manuscript) taken with a superconducting tip on the 5$^{th}$ atom from the top of the chain, together with the fits from numerical deconvolution of the spectra employing four peaks due to YSR states, in two different situations: (i) all the adatoms in the chain are in YSR-on state (top curves), (ii) 6$^{th}$ atom is in YSR-off state while all others are in YSR-on state (bottom curves). **b,** The numerically deconvoluted LDOS (same as Fig. 5f of the main manuscript) corresponding to the fits in panel **a**. The substrate DOS is plotted as a reference. **c,d**, similar plots as **a**, **b**, respectively, except that in this case only two peaks were employed for fitting the spectra. It is clear from the fits in **c** that the spectral shapes, especially at the positions indicated by black arrows, can be correctly captured only by employing four Lorentzian peaks (see black dash-dotted lines in **b**). This indicates a splitting of the YSR states and implies a spin coupling between the adatoms in the chain.

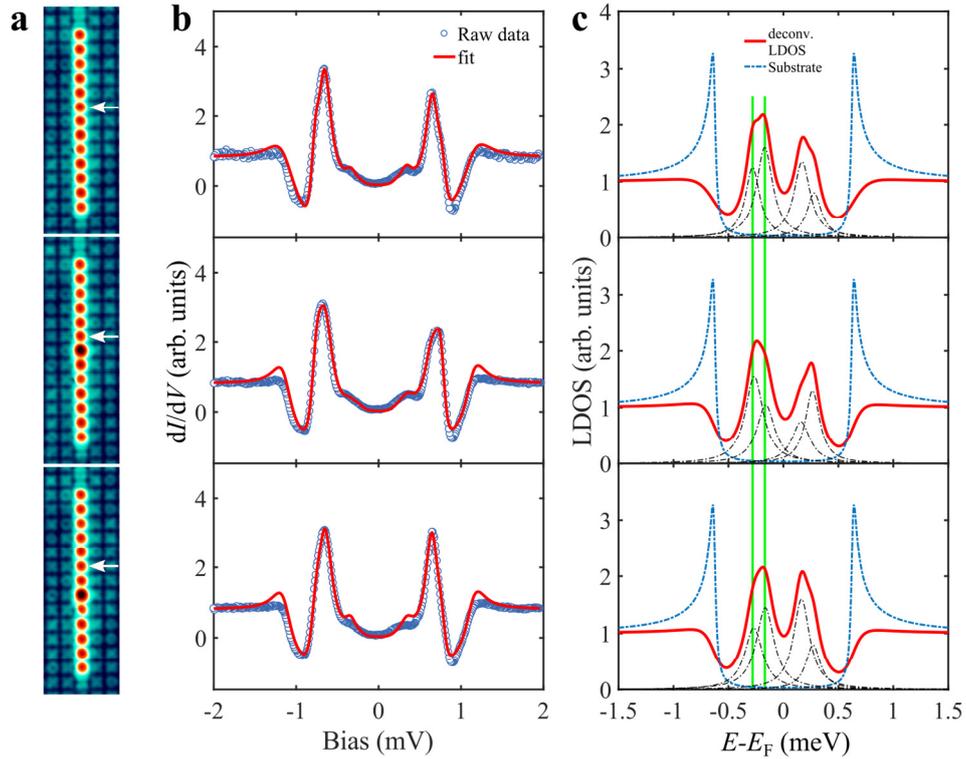

**Supplementary Figure 6 | Numerical deconvolution of the spectra on the atom in the chain. a**, Constant-current STM topographs of the chain (same chain as shown in Fig.5 of the main manuscript) for the three different situations: all the atoms are in YSR-on state (top), (ii) 7$^{th}$ atom from the top is in YSR-off state while rest of the atoms are in YSR-on state (middle panel), (iii) 8$^{th}$ atom from the top is in YSR-off state while rest of the atoms are in YSR-on state (bottom panel). **b**, d$I$/d$V$ spectra taken with a superconducting tip on the 6$^{th}$ atom from the top, together with the fits from the numerical deconvolution of the spectrum for the three cases shown in **a**. **c**, The numerically deconvoluted LDOS corresponding to the fits in each panel in **b**. The substrate density of states is plotted as a reference. Vertical green lines in **c** are guides to the eye showing the peak positions of the YSR states in the upper most panel. The deconvoluted LDOS clearly shows that at least four prominent peaks are needed in order to correctly capture the details of the spectra in **b** (see black dash-dotted lines). This indicates that there is a considerable spin coupling between the adatoms in the chain. From the vertical green lines in **c** it is evident, that we do not see significant changes in the separation between the split peaks when switching one of the nearest or next-nearest neighbors into the non-magnetic state. However, the intensity difference between the two peaks on the negative (or positive) bias side is inverted when the 7$^{th}$ atom in the chain is switched to the YSR-off state, and the intensities are regained when the 8$^{th}$ atom is switched to the YSR-off state. This observation is different than the case described in Figure 5**e,f** of the main manuscript, where upon switching the 6$^{th}$ atom from YSR-on to YSR-off state, the separation between the two split YSR states on the neighbouring 5$^{th}$ atom reduces. The different behavior of the split YSR states for the two atoms further confirms an overall inhomogeneous and non-trivial spin-coupling along the chain.

## Supplementary References: